\documentclass{bmcart}

\usepackage{amsthm,amsmath}
\usepackage[utf8]{inputenc} %unicode support

\usepackage{algorithm2e}
\usepackage{graphicx}
\usepackage{soul} %for striking through text
\usepackage[super]{nth} % for superscripts
\usepackage{comment}

\startlocaldefs
\endlocaldefs

\begin{document}

%%% Start of article front matter
\begin{frontmatter}

\begin{fmbox}
\dochead{Research}


\title{A network perspective on intermedia agenda-setting}

%%%%%%%%%%%%%%%%%%%%%%%%%%%%%%%%%%%%%%%%%%%%%%
%%                                          %%
%% Enter the authors here                   %%
%%                                          %%
%% Specify information, if available,       %%
%% in the form:                             %%
%%   <key>={<id1>,<id2>}                    %%
%%   <key>=                                 %%
%% Comment or delete the keys which are     %%
%% not used. Repeat \author command as much %%
%% as required.                             %%
%%                                          %%
%%%%%%%%%%%%%%%%%%%%%%%%%%%%%%%%%%%%%%%%%%%%%%

\author[
   addressref={aff1},                   % id's of addresses, e.g. {aff1,aff2}
                        % id of corresponding address, if any
%   noteref={n1},                        % id's of article notes, if any
   email={samuel.stern.16@cs.ucl.ac.uk}   % email address
]{\inits{SS}\fnm{Samuel} \snm{Stern}}
\author[
   addressref={aff1,aff2},
   corref={aff1},
   email={g.livan@ucl.ac.uk}
]{\inits{GL}\fnm{Giacomo} \snm{Livan}}
\author[
   addressref={aff1},
   email={g.livan@ucl.ac.uk}
]{\inits{RES}\fnm{Robert E.} \snm{Smith}}

%%%%%%%%%%%%%%%%%%%%%%%%%%%%%%%%%%%%%%%%%%%%%%
%%                                          %%
%% Enter the authors' addresses here        %%
%%                                          %%
%% Repeat \address commands as much as      %%
%% required.                                %%
%%                                          %%
%%%%%%%%%%%%%%%%%%%%%%%%%%%%%%%%%%%%%%%%%%%%%%

\address[id=aff1]{%                           % unique id
  \orgname{Department of Computer Science, University College London}, % university, etc
  \street{Gower Street},                     %
  \postcode{WC1E 6BT}                                % post or zip code
  \city{London},                              % city
  \cny{UK}                                    % country
}

\address[id=aff2]{%                           % unique id
  \orgname{Systemic Risk Centre, London School of Economics and Political Sciences}, % university, etc
  \street{Houghton Street},                     %
  \postcode{WC2A 2AE}                                % post or zip code
  \city{London},                              % city
  \cny{UK}                                    % country
}

%%%%%%%%%%%%%%%%%%%%%%%%%%%%%%%%%%%%%%%%%%%%%%
%%                                          %%
%% Enter short notes here                   %%
%%                                          %%
%% Short notes will be after addresses      %%
%% on first page.                           %%
%%                                          %%
%%%%%%%%%%%%%%%%%%%%%%%%%%%%%%%%%%%%%%%%%%%%%%

\begin{artnotes}
%\note{Sample of title note}     % note to the article
%\note[id=n1]{Equal contributor} % note, connected to author
\end{artnotes}

\end{fmbox}% comment this for two column layout

%%%%%%%%%%%%%%%%%%%%%%%%%%%%%%%%%%%%%%%%%%%%%%
%%                                          %%
%% The Abstract begins here                 %%
%%                                          %%
%% Please refer to the Instructions for     %%
%% authors on http://www.biomedcentral.com  %%
%% and include the section headings         %%
%% accordingly for your article type.       %%
%%                                          %%
%%%%%%%%%%%%%%%%%%%%%%%%%%%%%%%%%%%%%%%%%%%%%%

\begin{abstractbox}

\begin{abstract} % abstract
In Communication Theory, intermedia agenda-setting refers to the influence that different news sources may have on each other, and how this subsequently affects the breadth of information that is presented to the public. Several studies have attempted to quantify the impact of intermedia agenda-setting in specific countries or contexts, but a large-scale, data-driven investigation is still lacking. Here, we operationalise intermedia agenda-setting by putting forward a methodology to infer networks of influence between different news sources on a given topic, and apply it on a large dataset of news articles published by globally and locally prominent news organisations in 2016. We find influence to be significantly topic-dependent, with the same news sources acting as agenda-setters (i.e., central nodes) with respect to certain topics and as followers (i.e., peripheral nodes) with respect to others. At the same time, we find that the influence networks associated with most topics exhibit small world properties, which we find to play a significant role towards the overall diversity of sentiment expressed about the topic by the news sources in the network. In particular, we find clustering and density of influence networks to act as competing forces in this respect, with the former increasing and the latter reducing diversity.   
%This study operationalises intermedia agenda-setting within the media and examines the problem of empirically measuring and monitoring the formation of opinion in the online news media and the role that peers play in influencing how agents adopt opinions. To explore network opinion dissemination in the media, this paper proposes a technique for constructing social networks, based on Granger (non-)causality of published content, that capture how the media may be influenced to adopt opinions from their peers. We examine the characteristics of the social network structure of these relationships in the context of intermedia agenda-setting. Through comparing networks of influencer-influencee relationships we find that while there is no single agenda-setting entity, there are communities of densely connected agents. Furthermore, we consider whether the network characteristics that emerge in the media impact the level of diversity of opinions that the public is exposed to. Our results suggest that higher degrees of interaction within the media reduce the overall opinion diversity, however, this is reduced for issues in which the media form distinct clusters.
%\parttitle{First part title} %if any
%Text for this section.

%\parttitle{Second part title} %if any
%Text for this section.
\end{abstract}

%%%%%%%%%%%%%%%%%%%%%%%%%%%%%%%%%%%%%%%%%%%%%%
%%                                          %%
%% The keywords begin here                  %%
%%                                          %%
%% Put each keyword in separate \kwd{}.     %%
%%                                          %%
%%%%%%%%%%%%%%%%%%%%%%%%%%%%%%%%%%%%%%%%%%%%%%


% MSC classifications codes, if any
%\begin{keyword}[class=AMS]
%\kwd[Primary ]{}
%\kwd{}
%\kwd[; secondary ]{}
%\end{keyword}

\end{abstractbox}
%
%\end{fmbox}% uncomment this for twcolumn layout

\end{frontmatter}

%%%%%%%%%%%%%%%%%%%%%%%%%%%%%%%%%%%%%%%%%%%%%%
%%                                          %%
%% The Main Body begins here                %%
%%                                          %%
%% Please refer to the instructions for     %%
%% authors on:                              %%
%% http://www.biomedcentral.com/info/authors%%
%% and include the section headings         %%
%% accordingly for your article type.       %%
%%                                          %%
%% See the Results and Discussion section   %%
%% for details on how to create sub-sections%%
%%                                          %%
%% use \cite{...} to cite references        %%
%%  \cite{koon} and                         %%
%%  \cite{oreg,khar,zvai,xjon,schn,pond}    %%
%%  \nocite{smith,marg,hunn,advi,koha,mouse}%%
%%                                          %%
%%%%%%%%%%%%%%%%%%%%%%%%%%%%%%%%%%%%%%%%%%%%%%

%%%%%%%%%%%%%%%%%%%%%%%%% start of article main body
% <put your article body there>

%%%%%%%%%%%%%%%%
%% Background %%
%%

\section{Introduction}
\label{intro}
 
The news media is an important information source for much of the world's population, as it feeds into opinions and choices \cite{mccombs_agenda-setting_1972,mccombs_candidate_1997,wilczek_herd_2016}. Frequently referred to as the `gatekeepers of information', journalists have been shown to play a powerful role in influencing public opinion through determining what stories (or elements thereof) are presented to the public, how content is framed, which elements are emphasized, and how the public forms associations between topics covered by the news \cite{guo_expanded_2012,wilczek_herd_2016}. While much research has been done on identifying bias in the mainstream media \cite{wilczek_herd_2016}, and has more recently focused on `fake news' and misinformation (e.g., \cite{vargo_agenda-setting_2018,shu_beyond_2019,sikder2020minimalistic}), little research has thus far looked at studying the dynamics underpinning the formation of opinions within the media. News organisations do not construct narratives in isolation, but are instead influenced by each other. This phenomenon is the subject of \emph{intermedia agenda-setting theory} \cite{harder_intermedia_2017}, which explains the `co-orientation' of the narratives in the media as the result of exposure to both economic and local social factors \cite{guo_expanded_2012}. The topics that the media covers and the context in which headlines are presented have been shown to impact observable phenomena ranging from voter behaviour \cite{mccombs_candidate_1997} to macroeconomic indicators \cite{nyman_algorithmic_2016}. Affective attributes of the news content, such as sentiment and tone, have been shown to affect voters' judgments on political issues \cite{coleman_proposing_2010,sheafer_evaluate_2007} and people's perceptions of corporations \cite{kim_role_2012}. Because of its far-reaching impact, it is valuable to understand the opinion dynamics of the media at the level of individual news organisations as it ultimately determines what content is made available to the public as well as how it is presented.

The media continues to play a central role in directing what information the public attends to, how it is presented, and with what they associate that information. It is therefore important to have a stable, balanced and diverse set of opinions being presented to the public. Research on agenda-setting theories of the media \cite{mccombs_agenda-setting_1972} generally either aggregate the media into a single representative agent or treat the media as a collection of independent actors. Here, we demonstrate an alternative representation in which news sources can be understood as nodes in a social network whose actions are influenced by the other nodes in the network. By framing the system as a social network, we are able to exploit the tools provided by network science to aid in our understanding of how such a system evolves. Numerous studies have demonstrated how network structure impacts signal propagation \cite{delre_diffusion_2007,watts_influentials_2007}. However, we are not aware of any research to date that has employed a network approach to explore the propagation of information within the media. 

In many real-world social networks, connections and interactions between agents can be explicitly gathered from the data, such as through citation in academia \cite{li2019reciprocity}, or liking and following in social media \cite{anger_measuring_2011,livan2017excess}. In the domain of the news media, whilst there are naturally interactions between agents, these are latent and network structures must first be inferred from the available data before one can study the dissemination and propagation of news and sentiment through them. Previous work on studying the dissemination of news data from a social network modelling approach has explored what content the media presents \cite{vargo_networks_2017} and how breaking news stories are picked up and propagated through the media \cite{liu_breaking_2016}, but not how the network of interactions of the news sources impacts how content is framed.

In this work, we tackle the above by presenting a method to operationalise intermedia agenda-setting theory in which each news source is a potential node in a directed network and edges exist between news sources when there is empirical evidence that the sentiment expressed by one source on a topic is Granger-caused by that of another. We implement the proposed method and employ it to answer the following research questions (RQs):
\begin{itemize}
\item RQ1: Can `bellwether' news sources be identified? That is to say, sources who have influence over the way in which the rest of the media portray a story. Moreover, are such bellwethers more central in networks that capture intermedia influence?
\item RQ2: Is there emergent structure in the networks that can potentially impact how a news story is depicted and propagated across the media?
\item RQ3: How does the structure of these networks impact the diversity of sentiment that exists across the media on a given topic? 
\end{itemize}

We propose and implement the aforementioned method, which constitutes our first contribution. We perform topic modelling on a collection of news articles to identify 200 topics discussed in the news media between May and December 2016. Sentiment analysis is then used to model the views adopted by each news source and how these vary over time. Lead-lag relationships between the sentiment expressed across news sources are identified to construct networks representing how sentiment is transmitted and adopted by the media. Our second contribution is then in applying this method to explore the diffusion of sentiment in the media. Our final contribution is towards the discussion of the ability of the media to self-regulate and provide a varied perspective for public opinion formation. 

We find:
\begin{enumerate}
\item 	There exist significant lead-lag relationships in which specific news sources can influence the sentiment with which a story is presented. 
\item Intermedia influence networks are able to capture characteristics that are consistent with intermedia agenda-setting theory, such that certain elite news sources sit more centrally.
\item Influence is distributed across media sources and bellwether behaviour varies depending on the topic.  
\item A significant proportion of topics exhibit `gatekeeping' tendencies in which a small subset of news sources have disproportionately large influence.
\item Influence networks exhibit small-world characteristics with communities of densely connected subgroups.
\item The overall diversity in sentiment is impacted by the structural features of intermedia networks. In particular, the density of such intermedia networks has a negative impact on the sentiment diversity on a topic whereas the strength of the network clustering has a positive impact on it.
\end{enumerate}

The remainder of this paper is organised as follows: in section \ref{related_work} we provide a background and overview of related work on influence analysis in social networks. In section \ref{methods} we outline our methodology, and in section \ref{results} we provide a detailed analysis of our results and discuss their implications. Finally, section \ref{conclusion} provides some conclusive remarks and proposes potential further work. 

\section{Related Work}
\label{related_work}
Since the beginning of the \nth{20} century, numerous theories have emerged that try to explain the role of the media in society and the impact that it has on public opinion (and vice versa). Arguably the most broadly accepted theory today is based on \textit{agenda-setting theory} \cite{mccombs_agenda-setting_1972} and extensions thereof. It argues that the media does not regulate what audiences think, but rather it tells them what to think \textit{about} by controlling what content is accessible. It paints the media as `gatekeepers’ of information about reality by selecting, omitting, and framing what issues are reported on, and to what extent. This then prompts the public to perceive selected issues as being more/less important than others, or drawing links between events, which ultimately changes how the audience engages with the issue and thereby influences the public’s narrative \cite {mccombs_look_2005}.

Beyond looking at the relationship between the media and the public, \textit{intermedia agenda-setting} research concerns itself with measuring the extent to which news content transfers between media. The highly regarded media have been shown to exert influence over their peers \cite{vargo_networks_2017, harder_intermedia_2017}. This ``co-orientation'' within the media can be attributed to several factors. One explanation is that it is the result of `churnalism' whereby media sources re-hash the content of their peers without having to expend the resources that are required to construct the narrative \cite{harder_intermedia_2017, lim_first-level_2011}. Another theory is that co-orientation is sociopsychological in origin, whereby the major news sources are regarded as better judges of what is (subjectively) newsworthy \cite{harder_intermedia_2017,harcup_what_2017}. Recently intermedia agenda-setting has also considered the transfer of content between different types of media, such as whether, and when, the mainstream media influences social media, and vice versa. It has been shown, for instance, that the online news media influences the content of politically aligned Twitter accounts \cite{harder_intermedia_2017}. 

So far, intermedia agenda-setting research has focused on testing \textit{a priori} hypotheses about the expected influence exerted by certain news sources. For example \cite{vargo_networks_2017} find that \textit{The New York Times} and \textit{The Washington Post} have the ability to set the agenda of others. In \cite{welbers_gatekeeper_2018}, the authors find a similar effect for \textit{Algemeen Nederlands Persbureau (ANP)} in the Dutch press, while \cite{lim_first-level_2011} identifies intermedia agenda-setting effects of \textit{Chosun.com} and \textit{ Donga.com} on other South Korean online media websites. 

Here, by operationalising intermedia agenda-setting in network terms, we circumvent the need for \textit{a priori} assumptions about which sources may act as agenda-setters. A core component in this is to automatically capture and interpret how  the  agents  in  the  news  media  influence  one  another. As will be discussed in the next section, the concept of `influence', and how it is measured, has emerged as a popular field of research in applied mathematics and computer science.

\subsection{Social Network Influence}
There is no single mathematical definition of social network influence. This is largely because it varies between contexts, and even within the same domain, it can be unclear how influence is defined. Peng et al. \cite{peng_influence_2018} describe four basic components that define influence, regardless of the domain: (1) social influence exists between two agents, the influencer and the influencee (aka. the influenced); (2) influence is a function of uncertainty, with zero uncertainty if the influencer has no doubt that the influencee will perform an action and the maximum amount of uncertainty when the influencer has no influence over the influencee's actions; (3) the level of influence can be represented by a real number which describes the uncertainty of the influencer's ability to dictate the influencee's action; (4) influence does not need to be symmetric. 

Consistent with the above criteria, numerous evaluation metrics have been proposed to assign influence values to agents in social networks, some based purely on centrality and network topology, and others on entropy measures. Degree, closeness, PageRank, HITS and Katz centralities have all been used in past studies \cite{kiss_identification_2008, peng_influence_2018, romero_influence_2011, weng_twitterrank:_2010}.

While influence measures based on network topology are convenient because they may be calculated for an arbitrary network, they ignore potentially valuable information about the content of the signals being passed through networks. Entropy measures of influence are one method for including signal content. Ver Steeg and Galstyan \cite{ver_steeg_information-theoretic_2013}, for example, propose the use of \textit{transfer entropy}, a measure for determining how much better the sequence of signals emitted by one agent can be predicted by including the signal history from another agent. Other methods have focused on whether the signals transmitted by one agent Granger-cause those of the other agents within the same network \cite{vargo_networks_2017,stern_measuring_2018}. We will build networks of influence based on the latter approach.

\subsection{Opinion and Emotion Influence}

Prior research has approached the question of how actors influence one another through emotion from two differing disciplines. The idea of how emotions and opinions are influenced has been studied extensively in the fields of psychology, cognitive science and sociology \cite{hatfield_emotional_1993,moors_theories_2009,van_kleef_emotion_2011}. More recently though, computational and mathematical models have explored how sentiment and opinions spread through social networks. The addition of emotion in the context of influence within social networks has been examined in Twitter \cite{stieglitz_emotions_2013,ferrara_measuring_2015,xiong_emotional_2018}, Facebook \cite{coviello_detecting_2014}, and blog posts \cite{hui_quantifying_2010,song_identifying_2007}. Semantic information expressed by actors in social media has been proposed as a method to augment influencer detection beyond what is possible using only network topology \cite{song_identifying_2007,zhou_finding_2009}. Inversely, incorporating information about a user's influence based on network topology has been used to augment sentiment analysis \cite{eliacik_influential_2018,hui_quantifying_2010}. Furthermore, contagion models have been developed to study how emotion and emotive content propagates and diffuses throughout social networks; sentiment in social media-based content has been shown to correlate with both the speed and quantity of information sharing \cite{brady_emotion_2017, hill_emotions_2010, stieglitz_emotions_2013, xiong_emotional_2018}. 

\section{Methodology}
\label{methods}

\subsection{Identifying Sentiment-Loaded Topics}
Whereas prior related studies have looked at how news stories are picked up on, we are concerned not only with what is discussed but also with capturing how it is presented. In other words, understanding from what angle the story or `theme' is depicted as well as how this varies across different news sources and different topics. Therefore, our goal is to construct time-series that encapsulate two factors: (1) the prominence that a source assigns to a recurring topic, and (2) the sentiment with which it covers it. We do this by performing topic modelling on news articles to identify what the stories are about; we then scale the topic assignments according to the sentiment of the articles. This gives us an indicator of the affinity between the source of an articles and the topics the article covers.

Our method produces a $|T| \times |A| \times |K|$ tensor, where $|K|$ is the number of topics, $|A|$ is the number of news sources, and $|T|$ is the length of our time-series. In other words, each news source, $a \in A$ has $|K|$ distinct time-series that describe how their sentiment towards each of the topics evolves through time. Topic extraction is performed by training a Latent Dirichlet Allocation (LDA) topic model \cite{blei_latent_2003} with $|K|$ topics on a corpus containing news articles from a range of different sources. LDA is an unsupervised learning technique that models documents as a multinomial distribution over $|K|$ topics, and topics as a multinomial distribution over $|V|$ words. It has been shown to successfully capture semantically coherent clusters of topics in medium-length texts such as news articles, blogs and journals \cite{blei_latent_2003}. Using the LDA model, each article in the corpus is mapped to a multinomial topic distribution that encodes how `much' of each of the $|K|$ topics is entailed within the article. Topics are modelled as being static over time (i.e., the per-topic word distributions are independent of time) because long-term media coverage has been shown to have a greater effect on policy-making than topics that are covered over the short- or medium-term through repetition and cumulative exposure \cite{joly_disentangling_2016}.

The sentiment for each article is subsequently measured following \cite{tuckett_tracking_2014}, i.e., counting the relative frequency of 'excitement' and `anxiety' keywords that have been shown to capture the affective impact that articles may have on a reader. That is, the sentiment value of an article is the number of excitement keywords minus the number of anxiety keywords, all as a fraction of the length of the article. Despite there being more sophisticated methods of capturing the sentiment of a document, using pre-defined keywords allows us to explicitly remove those words from the vocabulary that constitutes the topic distribution of words in the LDA model. This ensures that when the sentiment dynamics of each of the topics are later analysed, they are not biased by intrinsically `positive' or `negative' topics. 

To capture how a particular news source `feels' about a topic on any given day, each of that source's documents, from that day, are assigned a topic distribution using the LDA model.
The topic probabilities for each of that source's articles are scaled by the sentiment score of that article and then summed together across articles. Repeating this process daily for each news source allows us to populate a tensor $G$ where $G_{t,a,k}$ is the sentiment of a news source $a$ on topic $k$ on a day $t$. The steps for assigning topic-sentiment time series are presented in algorithm \ref{algo: TSA}. 

\begin{algorithm}
\SetAlgoLined
%\setstretch{1.1}
\SetKwInOut{Input}{input}
\SetKwInOut{Output}{output}
\Input{Set of articles $D$}
\Output{Tensor $G$ of size $|T| \times |A| \times |K|$}
$T \longleftarrow$ dates of articles in $D$\;
$K \longleftarrow$ set of topics\;
$A \longleftarrow$ set of news sources\;
\For{$t \in T$}{
	\For{$d \in D_t$}{
		$a \longleftarrow$ source of document $d$\;
		$\phi_d \longleftarrow$ multinomial topic distribution of document $d$\;
		$s_d \longleftarrow$ sentiment of document $d$\;
		$\psi_d \longleftarrow s_d \times \phi_d$\;
		\For{ $\psi_{k,d} \in \psi_d$}{
			$G_{t,a,k} \longleftarrow G_{t,a,k} + \psi_{k,d}$	
		}
	}
}
 \caption{Topic-Sentiment Assignment}
 \label{algo: TSA}
\end{algorithm}

The sentiment score assigned to a given news source on a given topic on a given day is influenced by three factors: the strength of the emotion in the content they published, the topic entropy of their articles (i.e., the probability mass of the articles that are attributed to a topic), and amount that they invest in a topic (i.e., the number of articles that have non-zero probability mass assigned to the topic). In other words, a source's score will be high if there is substantial positive emotion expressed in their content and/or their writing is about a specific topic and/or there is a proportionately large number of their articles about said topic. Concerning the final point, we note that news sources who publish more frequently may be assigned greater absolute sentiment scores on a topic. Our analyses throughout the rest of the paper, therefore, will only hinge on the relative fluctuations of such scores, and will be agnostic to absolute size.

\subsection{Topic-Level Influence Networks}
\label{tlin}

There have been many proposed definitions of influence, and metrics vary widely depending on the discipline, application area and data. The definition used here follows that of \cite{vargo_networks_2017,stern_measuring_2018} in which, for two agents $A$ and $B$, $A$ is considered to influence the opinion of $B$ on a given topic if the prior opinion expressed by $A$ on that topic increases our ability to predict that of B. More formally, if the conditional probability of observing $B(t)$ is dependent on $A(t-\delta(t))$ for some time $t$ and lag $\delta(t)$, then $A$ influences $B$. This lead-lag relationship is measured using the Granger causality test \cite{granger_investigating_1969}. Source $A$'s sentiment leads that of $B$ on a particular topic if $A$'s sentiment time series Granger-causes that of $B$ on that topic with a $p$-value lower than 0.05. We account for family wise error rate by applying the false discovery rate  correction, which we implement with the Benjamini-Hochberg procedure \cite{benjamini_controlling_1995}.

\begin{figure}[ht]
  \centering
  \includegraphics[width=0.9\linewidth]{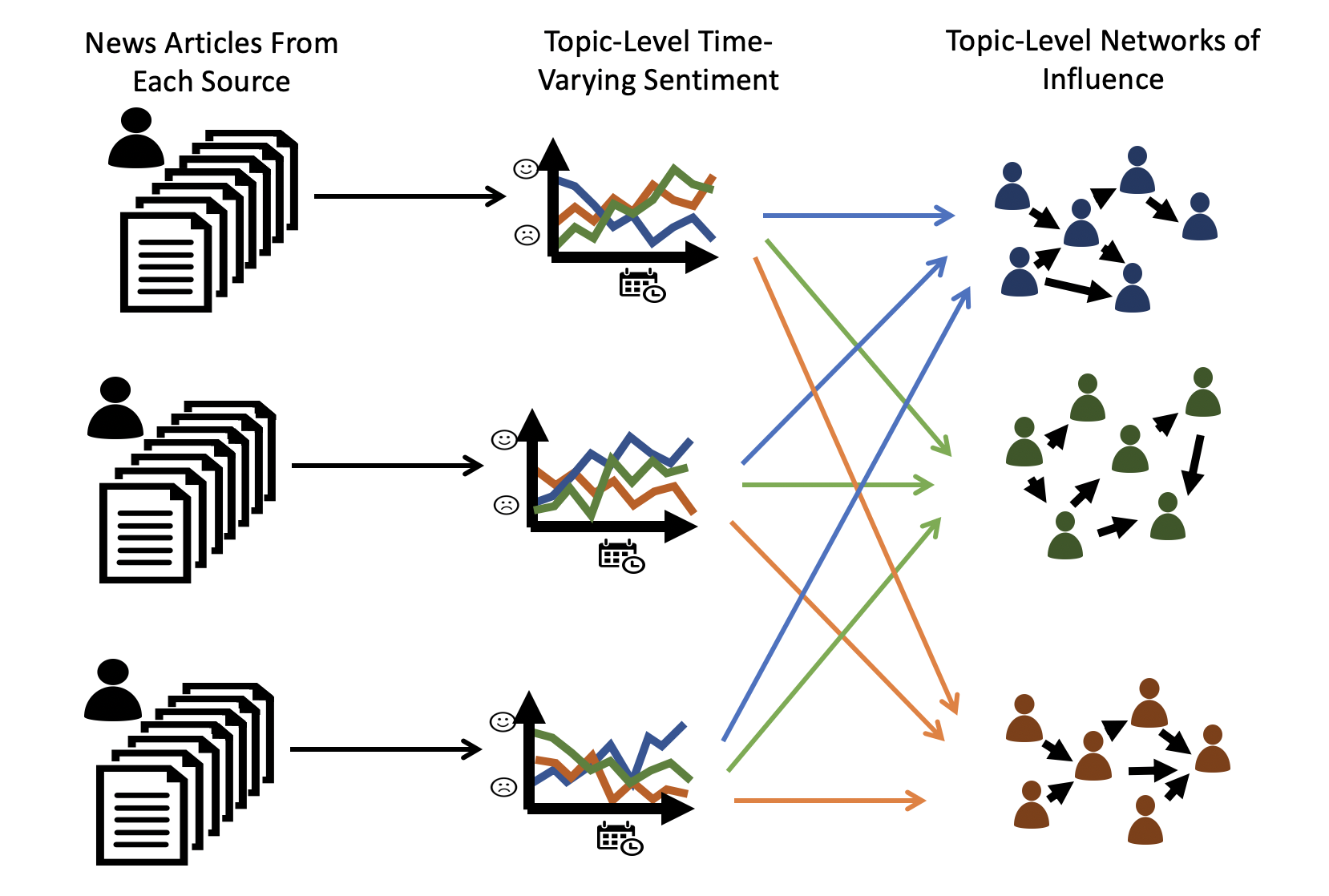}
  \caption{The network-construction process: A topic-specific sentiment time-series is inferred for each news source based on the content that they publish. Networks are inferred for each topic based on the empirically measured conditional dependence between the sentiment time-series of the news sources.}
  \label{diagram}
\end{figure}

For each pair of news sources, once lead-lag relationships between agents are identified, they can be used to populate topic-level influence networks. Graphs depicting the relationships between pairs of news sources are then constructed separately for each topic. Where a lead-lag relationship has been determined, both agents are added to the relevant graphs as nodes. An edge is drawn between them, with an arrow leading from the influencer to the influencee, resulting in an unweighted directed graph. This procedure is sketched in Figure \ref{diagram}.

\begin{figure}[ht]
  \centering
  \includegraphics[width=0.9\linewidth]{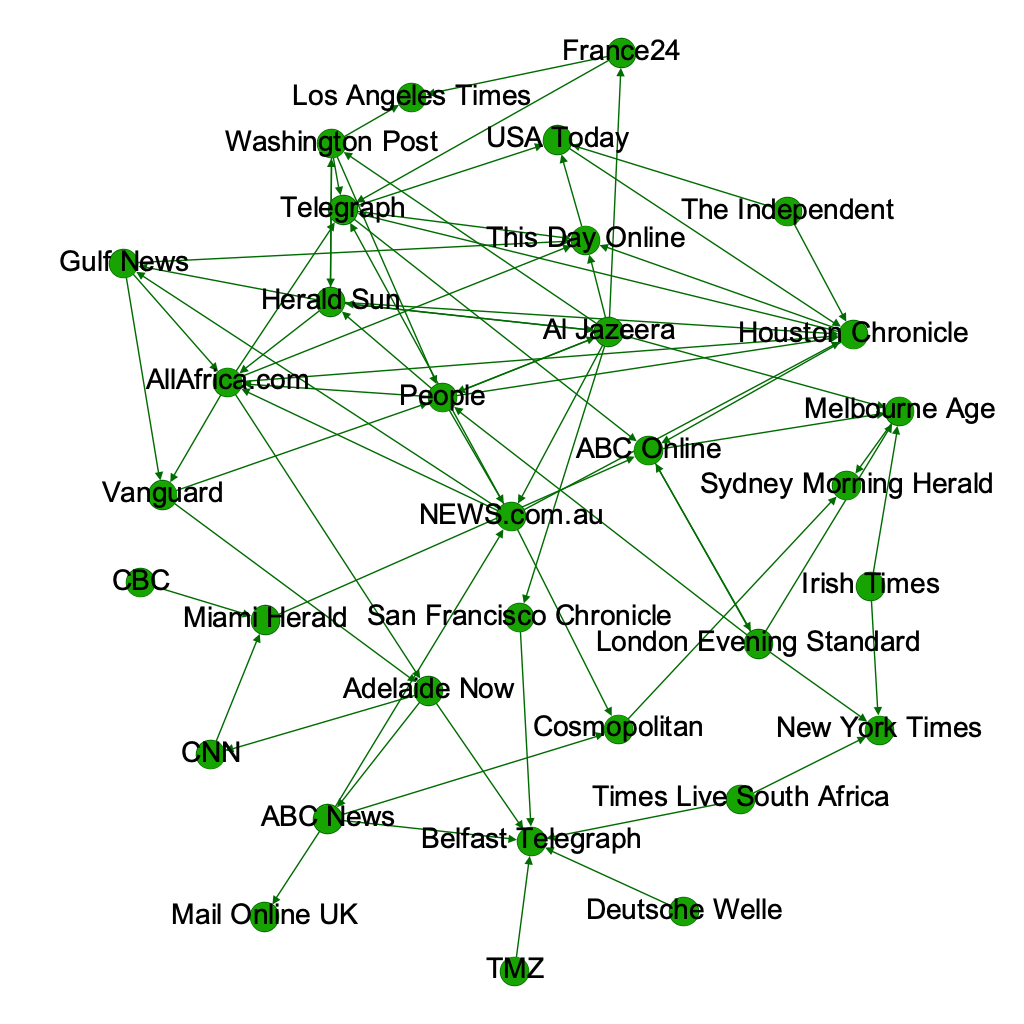}
  \caption{An example of the intermedia influence networks for a topic about the 2016 US presidential election. Each node is an agent within the news media and an edge exists going from an influencer to an influencee.}
  \label{topic_graph}
\end{figure}

Network sizes vary across topics because a news source is only added to a network if it has at least one edge. This results in sources that are not active in the discussion of a given topic, or which have no impact on their peers, to be excluded. Figure \ref{topic_graph} shows an example of such a network.

\subsection{Opinion Diversity}

One of the questions we wish to address is whether we can deduce how diverse the sentiment surrounding a topic is, given what we are able to capture of the influence that news sources have on one another. We, therefore, desire a measurable statistic that represents how varied news sources are in their sentiment towards a given topic. For this purpose, we define a measure of `opinion diversity' on a topic $k$ as the median of the daily cross-sectional variance of sentiment across news sources:

\begin{align}
\label{eq:op_div}
y_k &= \mathrm{median}\left\{\mathrm{Var}[G_{t,\bullet,k}]| t \in \{0,\dots,T\} \right \} \\ 
&= \mathrm{median} \left\{\frac{1}{|A|}\sum_{a}^{|A|}(G_{t,a,k}-\overline{G_{t,\bullet,k}})^2| t \in \{0,\dots,T\} \right \} \ .
\end{align}
Low values of $y$ indicate a general consensus in sentiment, whereas higher values reflect greater heterogeneity. We choose the median as a measure of central tendency as it is robust to skewed distributions, which in our domain arise when a topic receives a burst of news activity, such as a breaking news event.

%as the amount of variability in opinions, or degree of uncertainty, in the consensus on a topic. If there is, on average, low confidence in the consensus on a topic, then we argue that this indicates a high diversity in the opinions that are being expressed. Alternatively, high confidence in the consensus suggests that the various sources are in general agreeable. To capture this quantity for a given topic, we first take the cross-sectional variance at each point in time. We then measure the central tendency of this resulting series. We choose the median as a measure of central tendency as it is robust to skewed distributions, which in our domain arise when a topic receives a burst of news activity, such as a breaking news event.  Formally, the opinion diversity $y_k$ for the $k^{\mathrm{th}}$ topic is given as:

\section{Results}
\label{results}
The source of the news articles used in this research is the LexisNexis Newsdesk dataset. The dataset consists of 113,000,000 articles from over 34,000 news sources dating from 6 May 2016 to 31 Dec 2016. In addition to the content of the news articles, the dataset contains further metadata, including the source's country and broad topic labels such as `Top Stories', `Sport', and `Fashion'. To only focus on relevant articles, only those with between 100 and 1700 words in length and tagged under `Top Stories' are kept. Furthermore, to eliminate sources that only sporadically publish relevant content, all sources with an average publication rate lower than once per day are ignored. We also remove news aggregators, such as \textit{Yahoo! News}, and local subsidiaries of large national sources. This leaves 313,276 articles from 97 news sources usable in this study.   

Standard pre-processing is performed on each of the texts including tokenisation, lemmatisation and the removal of stopwords, before performing topic modelling and sentiment extraction on the articles.

Article topics are assigned using an LDA model of $k=200$ topics trained using the Gensim Python package \cite{rahman_hidden_2016}. The number of topics was chosen to maximise the model's coherence \cite{mimno_optimizing_2011} and a subset of the topics are presented in appendix A. Qualitative inspection of the topics learned by the LDA model suggests that the model successfully learns to cluster words based on identifiable news topics. These topics include Britain's exit from the European Union, which assigns high probability to words including [`europe', `brexit', `farage'], the scandal involving the state-sponsored doping of Russian athletes [`olympic', `doping', `russia'] and changes to central banks' interest rates [`bank', `rate', `bond'].

\subsection{Bellwether Behaviour}

\begin{figure*}[htb!]
  \centering
  \includegraphics[width=0.9\linewidth, height=6cm]{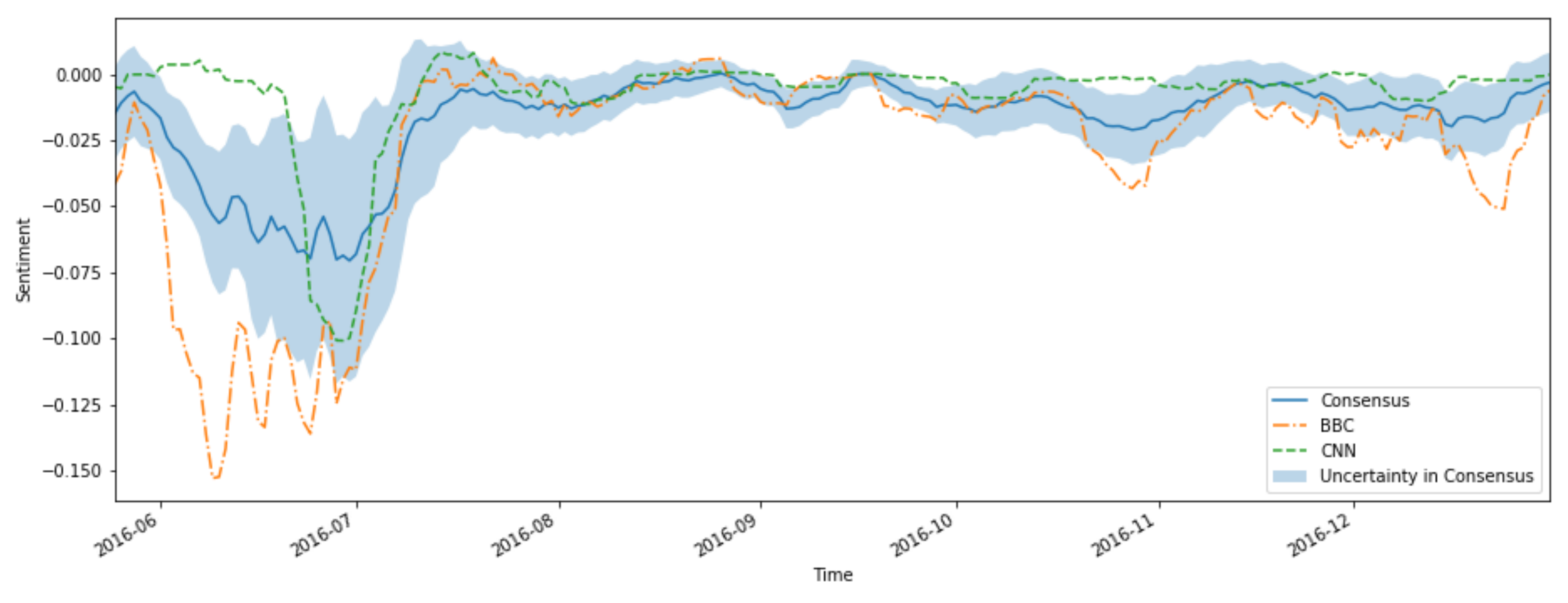}
  \caption{An example of how topic-specific sentiment changes over time. The example shown here is the `Brexit' topic. The BBC (orange) is a bellwether on this topic as it leads the consensus (blue) whereas CNN (green) lags the consensus.}
  \label{time-series}
\end{figure*}

Here we present the results and discussion concerning RQ1, which asks whether the sentiment that news sources adopt is influenced by those of their peers. 

There has been extensive research into herd behaviour and interdependence within the fields of behavioural economics and social psychology, in which studies have shown that an individual's bias towards a particular group is a significant contributor towards their social relations \cite{hogg_intergroup_2013}. This behaviour has also been proposed as a factor in determining how stories are framed in the media \cite{wilczek_herd_2016}. 

We distinguish between two indications of potential peer influence. The first `one-versus-all' approach considers whether the dynamics of a news source's sentiment on a particular topic influences, or is influenced, by the average sentiment expressed by the remainder of their peers and competitors, which we will refer to as the consensus. The second is whether there is a measurable relationship between distinct pairs of agents that cannot be explained away by the consensus. 

%The consensus view on a particular topic on any given day is measured by taking the mean of the topic-sentiment assignment across each of the news sources. This measure indicates the average of the collective sentiment at any point in time. 

To test whether an individual news source acts as a leading indicator on that topic it is first removed from the consensus and then tested to see whether the time-series that represents the consensus on a topic is Granger-caused by the news source in question. The inverse is also performed to determine if the sentiment of the source are Granger-caused by the consensus. This test is performed for each topic and each news source that contributes to that topic. We refer to sources that act as leading indicators as `bellwethers' and an example of this is presented in figure \ref{time-series}.

\begin{figure}[htb!]
  \centering
  \includegraphics[width=0.9\linewidth]{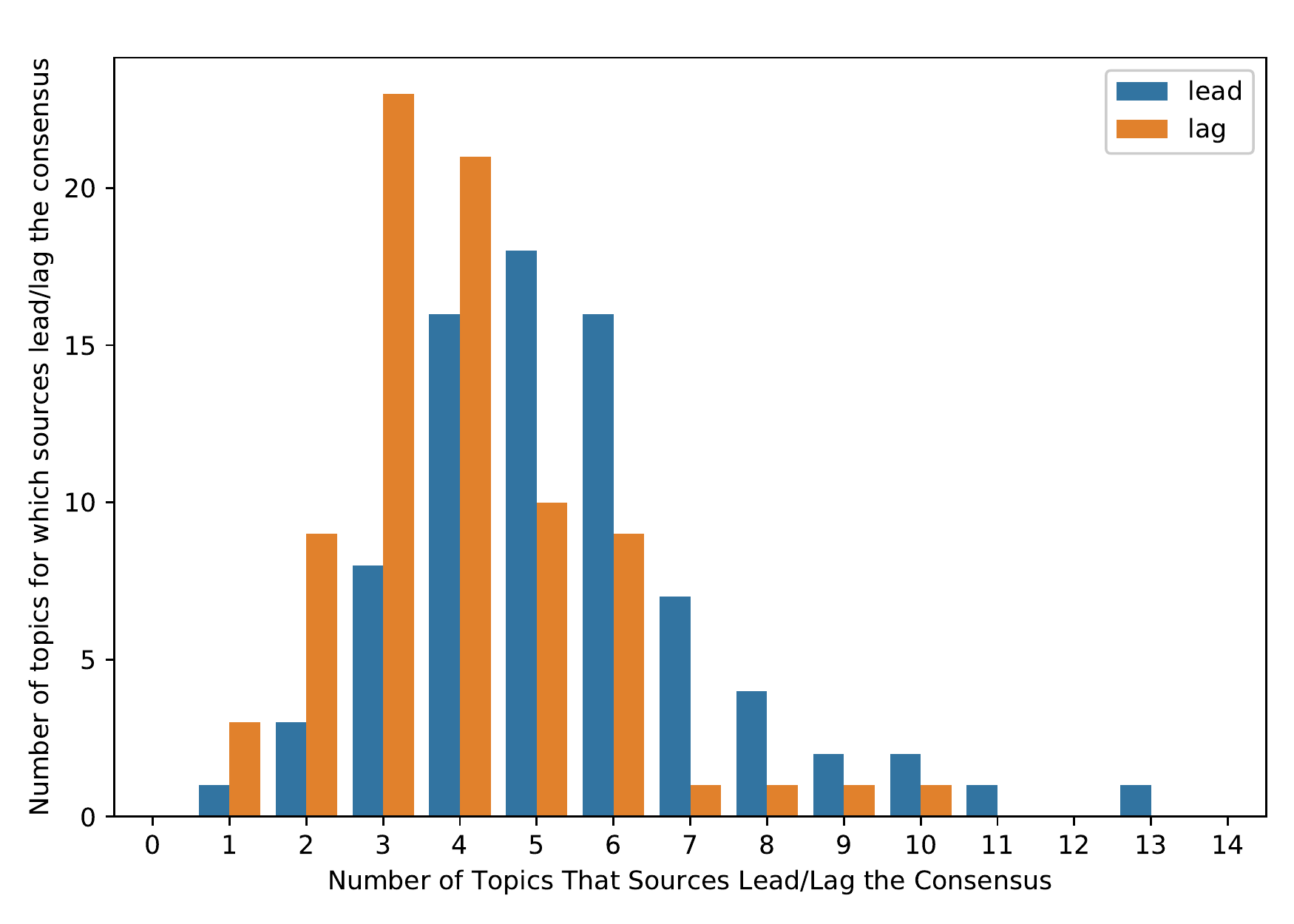}
  \caption{Histogram of the number of topics that each news source lead and lag. }
  \label{lead_lag_dist}
\end{figure}

Figure \ref{lead_lag_dist} presents an overview of the relative frequencies at which a news source leads versus lags the consensus. Of the 15,800 possible lead-lag relationships, there are 379 (2.4\%) cases in which a source Granger-causes the consensus and 256 (1.6\%) instances where a news source is Granger-caused by the consensus. 

Each of the 97 news sources acts as bellwether for at least one topic with the average news source being credited as a bellwether for 5 different topics. A manual inspection of the set of agents suggests that regional and national-level outlets are more likely to lead on topics within their own geopolitical environments than they are on non-local topics. This is conceivably because they are attuned to issues that are of concern to their readership base. Examples of this are the \textit{Miami Herald} being a leading source on topics surrounding the `Orlando Nightclub Shooting', or the \textit{London Evening Standard} being a leading indicator on the topic concerning `London', `Londoners' and `Sadiq Kahn'. Conversely, a minority of media outlets are leaders of a proportionately large number of topics. These are found to generally be larger outlets, such as the \textit{New York Times} and \textit{The Telegraph}, both of whom act as bellwethers on 13 topics, \textit{The Los Angeles Times} with 10 topics, and \textit{The Washington Post} with 9. Considering the instances where a source lags the consensus, the results also indicate that no single news outlet is wholly independent of other sources, but rather all have at least one topic for which they are trend followers. 

\begin{figure}[htb!]
  \centering
  \includegraphics[width=0.9\textwidth]{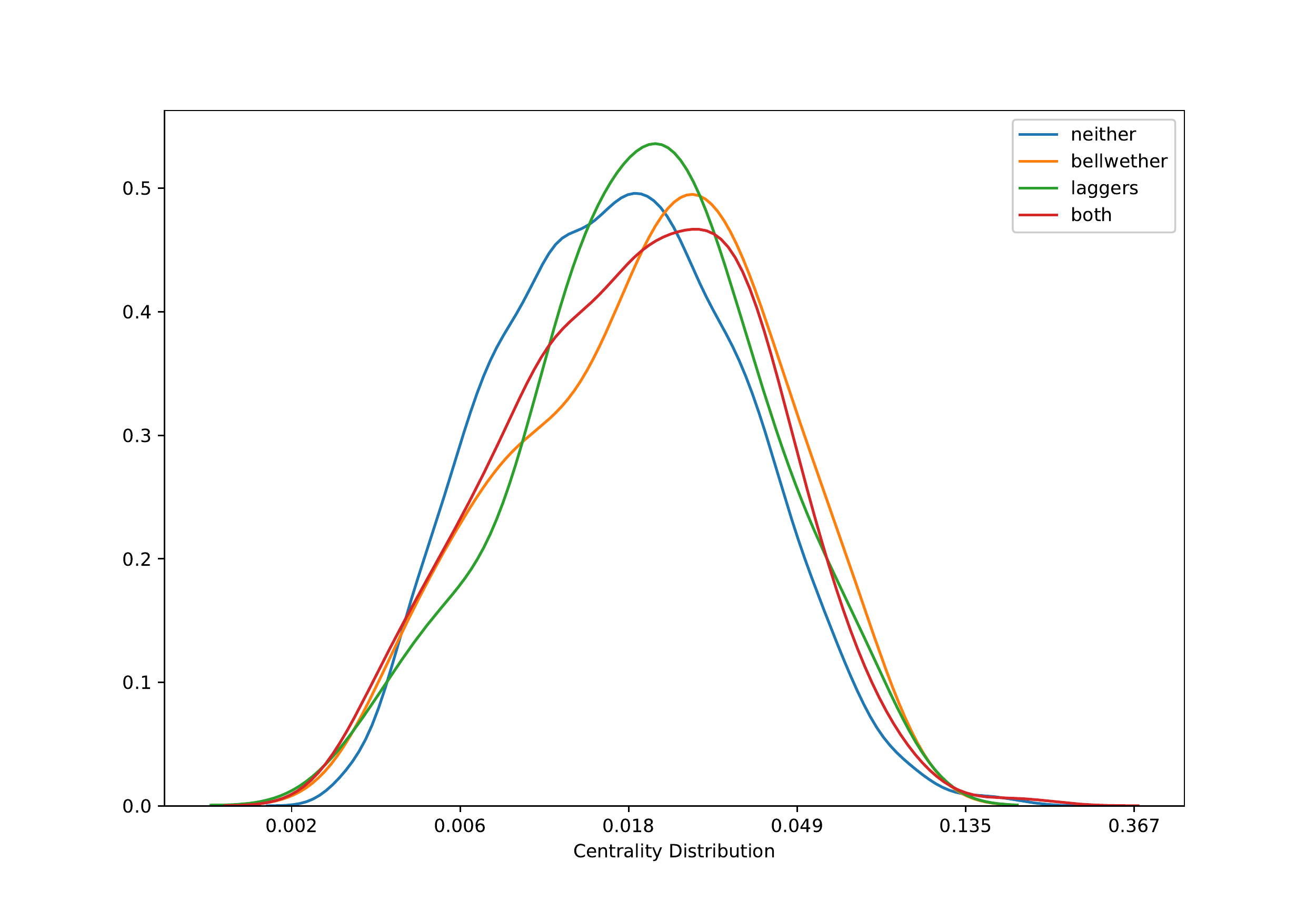}
  \caption{The distribution of PageRank centrality values grouped by whether a node is a bellwether, whether it lags the discussion, whether it neither leads nor lags the discussion or whether it both leads and lags the discussion.}
  \label{fig:lead_lag_cent_kde}
  
\end{figure}

This first result supports the notion that there is not only some level of interdependence in the topics which the media chooses to report on but also in the manner in which they do so. Next, we consider influence at a pairwise level and whether this phenomenon is mirrored in the influence networks, described in section \ref{tlin}. In particular, we test whether the bellwether news sources are more central in intermedia influence networks than those who are not bellwethers. For this, we assign each news source to one of four groups depending on whether it leads the consensus of a topic (i.e., a bellwether), whether it lags the consensus, whether it neither leads nor lags the consensus or whether it both leads and lags the consensus (mutual Granger causality). The PageRank centrality of each news source is calculated for each of the topic-influence networks that it sits in\footnote{The direction of the edges of each graph are reversed to reflect that PageRank assigns centrality to nodes whose incoming edges have high centrality}. The distributions of the resulting centrality measures of the groups, shown in figure \ref{fig:lead_lag_cent_kde}, are then compared by performing a Kruskal-Wallis H-test (KW test) for independent samples\footnote{The assumptions for an ANOVA are not satisfied}, which is a non-parametric test of whether two or more samples originate from the same distribution. More specifically, the null hypothesis of the KW test is that the medians of the samples are the same. We use the KW test to test whether the median centrality of a news source on a topic differs depending on whether that source was earlier identified as leading the topic consensus, lagging it, neither, or both. 

The KW test is performed jointly on the four groups and the results reject the null hypothesis that the groups are generated from the same distribution ($p=1.7 \times 10^{-4}$). While from this result we can assume that there is at least one group whose distribution stochastically dominates the others, it does not state which one. To determine where the stochastic dominance occurs we furthermore perform the test between pairs of groups, the results of which are shown in Table \ref{table:kw_results}. For any pair of groups, if the null hypothesis is rejected, then we can simply check which of the two groups has the larger median. The results table shows us that when a news source is a bellwether on a topic, it is also more likely to have higher centrality in the network associated with that topic than when it is not, but there is no statistical difference between those who lag the consensus and those who are neither leaders nor laggers.

\begin{table}[htb!]
\caption{The results of the Kruskal-Wallis H-test performed for each pair of groups. Adjusting for multiple comparison, the 10\%, 5\% and 1\% significance thresholds are 0.016, 0.008 and 0.0016 respectively}
\begin{tabular}{l|llll}
 & Lead & Lag & Both & Neither \\ \hline
Lead ($n=137$)& 1 &  &  &  \\
Lag ($n=102$)& 0.621 & 1 &  &  \\
Both ($n=605$)& 0.337 & 0.721 & 1 &  \\
Neither ($n=7826$)& 0.007** & 0.058 & 0.001*** & 1
\end{tabular}
\label{table:kw_results}
\end{table}

\subsection{The Structural Properties of Intermedia Influence Networks}

Having established that there is evidence of intermedia agenda-setting in which news sources are responsive to the content published by one another, we proceed by considering the structural properties of the networks that capture the intermedia influence.

\subsubsection{Intermedia Gatekeeping}

\begin{figure}[htb!]
  \centering
  %\begin{minipage}{.45\textwidth}
  \includegraphics[width=0.9\linewidth]{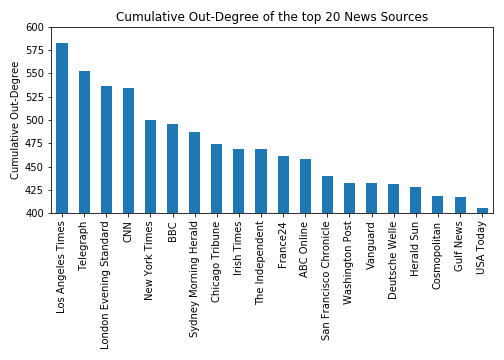}
  \caption{The cumulative out degree of the top 20 topics}
  \label{cumulative_out_deg}
  %\end{minipage}
  %\begin{minipage}{.45\textwidth}
  %\includegraphics[width=\linewidth]{gatekeeping_effect.png}
  %\caption{The percentage of topics who reject the null %hypothesis of $P(k \geq k_{max-x})$ for varying $x$}
  %\label{gatekeeping_effect}
  %\end{minipage}
\end{figure}

The `gatekeeping' function of the media is well documented in which, even in the age of social media, the editors act as information filters and barriers to the content that is disseminated to the public \cite{shoemaker_media_2014}. What is less well understood is whether there is intermedia gatekeeping in which a small subset of actors have influence that spans disproportionately far. We employ our inferred networks to address this question by testing whether individual news sources have disproportionately large influence that could enable them to govern the agenda of the discussion on particular issues. The scope here is not to definitively state whether or not certain players dominate the sentiment on a topic, but rather to ask whether the network structure facilitates this behaviour. 
We reason that if the media does have highly influential gatekeepers who can directly impact others, then we would expect to see a considerable proportion of topics where the associated network has at least one source that is significantly more central than would be expected to find if influence were assigned randomly. First we consider consider descriptive statistics about the aggregate influence. As seen in figure \ref{cumulative_out_deg}, we find that the larger news sources, such as \textit{The Los Angeles Times}, \textit{The Telegraph}, \textit{CNN} and \textit{BBC}, in general have greater total out-degree across networks. This qualitatively shows that influence is not uniformly distributed. However, from a quantitative standpoint it does not suggest whether this indicates disproportionately influential actors. We, therefore, run the following binomial test: for each network, $g \in G$, we begin by considering the out-degree of the most central node, $k_\mathrm{max}(g)$. We then ask what the probability would be of observing a node with out-degree $k\geq k_\mathrm{max}(g)$ under an Erd\H os-R\'enyi network generation process that has the same average degree as $g$. For a network $g$ with $n$ nodes and $np$ edges, if the probability, $P(k\ge k_\mathrm{max}(g))$, as given in equation \ref{eq:binom_cdf}, is smaller than a threshold $\alpha$ (where $\alpha$ includes Bonferroni correction), then we can argue that $g$ has at least one disproportionately central actor. We repeat this procedure for the second most central node, ($k_\mathrm{max-1}(g)$), third most central ($k_\mathrm{max-2}(g)$), etc.

\begin{equation}
\label{eq:binom_cdf}
P(k\ge k_\mathrm{max}(g)) = 1 - \sum_{i=1}^{k_\mathrm{max}(g)} \binom{i}{n} p^i (1-p)^{n-k}
\end{equation}

\begin{figure}[htb!]
  \centering
  \includegraphics[width=0.9\linewidth]{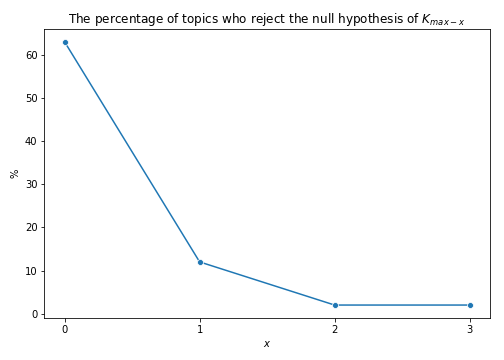}
  \caption{The percentage of topics who reject the null hypothesis of $P(k \geq k_{max-x})$ for varying $x$}
  \label{gatekeeping_effect}
\end{figure}

The results, seen in figure \ref{gatekeeping_effect} show that, after performing this test on each of the 200 topics, we observe 63 cases (31.5\%) where we can reject the null hypothesis that the centrality of the most central node in the network could be generated from an Erd\H os-R\'enyi process. From this result it is not possible to definitively conclude that there is systematic gatekeeping in the media. However, it does provide strong evidence that there are certain topics for which a small subset of the media hold a disproportionate amount of influence. We sought to corroborate this by performing a rich-club analysis~\cite{colizza_detecting_2006} (the rich-club effect is a widely known measure of the excess of interaction between nodes with a large degree in a network). However, the results were inconsistent and only a small minority of topics showed any evidence of the rich-club effect. This is to be expected given that connectivity in the domain of intermedia influence reasonably tends to be acyclic. From this we can conclude that for the topics that do exhibit gatekeeping tendencies, the influence appears to be enjoyed by only one or two sources who, however, are largely insulated from other influential sources.

\subsubsection{Small-World Properties}

The small-world property of many empirical social networks has been shown to have a profound impact on how efficiently information spreads through them \cite{latora_efficient_2001}. Due to their high clustering and small characteristic path length \cite{watts_collective_1998}, the nodes of networks that exhibit the small-world property locally sit within a close-knit community, whilst simultaneously still allowing for information to pass efficiently through the network as a whole. In networks that exhibit the small-world phenomenon, innovations (i.e., novel perspectives) by an individual are likely to spread through the network and be picked up by others. In the context of intermedia influence networks, this phenomenon translates to most sources appearing to be independent of one another, while in fact requiring only few steps for a new spin on a story to propagate through the network as a whole. 

There have been several proposed measures of small-worldness, each of which relies on a slightly varying interpretation of how a small-world network is defined \cite{humphries_network_2008,telesford_ubiquity_2011,neal_making_2015}. One popular method, proposed by Humphries and Davies \cite{humphries_network_2008}, measures small-worldness using $\sigma = \frac{\frac{C}{C_r}}{\frac{L}{L_r}}$, where $C$ and $L$ are the clustering coefficient and average path length of the network respectively, and $C_r$ and $L_r$ are the clustering coefficient and average path length of an equivalent random network. Though having been criticised as overestimating small-worldness on large networks \cite{telesford_ubiquity_2011},  $\sigma$ has the benefit of having a simple heuristic for classifying a network as small-world: if $\sigma > 1$ then the network is said to be small-world. The networks we consider are not particularly large (ranging from between 20 to 50 nodes per network), making the $\sigma$ value an appropriate choice. We use the definition proposed in \cite{fagiolo_critical_2007} that extends the definition of clustering from the undirected to directed graph case:

\begin{equation*}
C = \frac{1}{N}\sum_{i \in G} \frac{\frac{1}{2}\sum\limits_{j,h \in G}(e_{ij}+e_{ji})(e_{ih}+e_{hi})(e_{jh}+e_{hj})}{(K^{out}_{i} + K^{in}_i)(K^{out}_{i} + K^{in}_i-1)-2\sum\limits_{j \in G}e_{ij}e_{ji}} \ ,
\end{equation*} 
where $N$ is the number of nodes in the network, $e_{nm}$ is an an indicator function of whether there is an edge going from node $n$ to node $m$, and $K^{in}_n$ and $K^{out}_n$ are the in-degree and out-degree of node $n$ respectively. The average shortest path between directed nodes is defined as \cite{rubinov_complex_2010}: 

\begin{equation*}
L = \frac{1}{N(N-1)}\sum\limits_{j\in G, i \neq j} d_{ij} \ ,
\end{equation*}
where $d_{nm}$ is the shortest directed path length from node $n$ to node $m$.

To generate equivalent random networks that maintain equivalent degree sequences, we follow previous studies \cite{liao_small-world_2011} by randomly reshuffling edges, whilst preserving the in-degree and out-degree for each node.

\begin{figure}[htb!]
  \centering
  \includegraphics[width=0.9\linewidth]{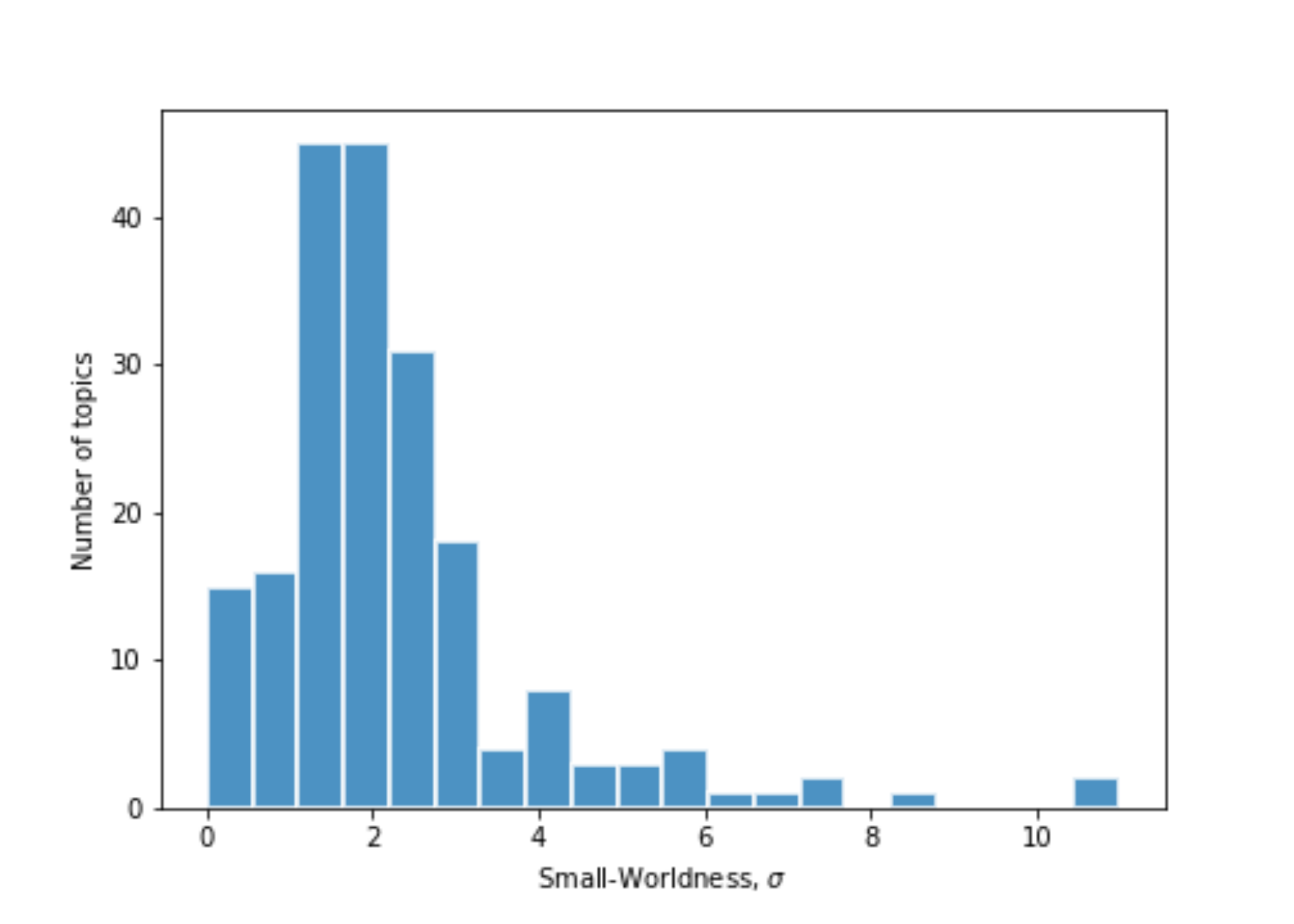}
  \caption{Histogram showing the distribution of small-worldness, $\sigma$. A value of $\sigma<1$ is considered not to be small-world.}
  \label{sigma_hist}
\end{figure}

Figure \ref{sigma_hist} shows a histogram of the  $\sigma$ values of the topic networks. 87.5\% of the networks have a value of $\sigma>1$, a strong indication that the media exhibits small-world tendencies in the manner in which they influence one another to adopt a particular angle or spin on a story.

\subsubsection{Opinion Diversity}

Now that we have a comprehensive view of the topology of intermedia influence networks, we proceed to explore the impact that the topology has on the level of opinion diversity observed across the media. 

Having a variety of opinions, emotions and viewpoints expressed by differing sources in the media is imperative to giving the public an opportunity to gain exposure to alternative perspectives on a particular issue \cite{mutz_facilitating_2001}. The above analysis of the structure of the networks indicates that there are both instances of identifiable community structures throughout the networks (which may isolate news sources from one another), as well as relatively short average distances between nodes (which conversely, may result in new information or perspectives spreading and being adopted through the network efficiently). In this final section, we seek to quantify the net impact that these two competing forces have on the diversity of sentiment expressed in the news media. To test this we propose a series of linear models that attempt to capture these effects and compare them based on the significance of their coefficients, while also controlling for the size and density of the networks.

We compare three linear regression models and examine their coefficients. The dependent variable, $\log(y)$, is the log of the opinion diversity, defined in Eq. \eqref{eq:op_div}. Each of the models contains as independent variables the average shortest path length $L$ as well as the size of the network $N$ (measured as the number of nodes) as a control variable. The first model, $M_1$, also includes the average clustering coefficient, $C$, as an independent variable while the second model, $M_2$, includes the network density, $d$, instead. The third model, $M_3$, considers the combined effect that both $d$ and $C$ have on $\log(y)$. 

Some topics covered in the news media may be intrinsically highly emotive, though not necessarily be particularly divisive or material to overall public opinion. Topics such as sport or celebrity gossip fall into this category, compared to, for example, media coverage about election candidates or international affairs. Therefore, the topics are inspected manually by the authors, and all of the topics that are clearly about sports, celebrity gossip or lifestyle/entertainment are ignored. This results in a sample size of 161 networks.

\begin{table}[h!]
\caption{The regression coefficients of each of the three linear models. Model $M_1$ is $\log(y) = \omega_0 + \omega_1 L + \omega_2 N + \omega_3 C $, model $M_2$ is $\log(y) = \omega_0 + \omega_1 L + \omega_2 N + \omega_4 d$, model $M_3$ is $\log(y) = \omega_0 + \omega_1 L + \omega_2 N + \omega_3 C + \omega_4 d$, where $y$ is the opinion diversity, $L$ is the average shortest path length, $N$ is network size, $C$ is the average  clustering coefficient, $d$ is the network density. The values shown below pertain to the coefficients $\omega_i$ ($i = 0, \ldots, 4$) with p-values reported in brackets. ($^* p < 0.1; ^{**} p < 0.05; ^{***} p < 0.01$.)}
\begin{tabular}{lccc}
\hline
\textbf{} &
  \textbf{$M_1$} &
  \textbf{$M_2$} &
  \textbf{$M_3$} \\ \hline
\textbf{Intercept} &
  \begin{tabular}[c]{@{}c@{}}-15.15\\ (0.001***)\end{tabular} &
  \begin{tabular}[c]{@{}c@{}}-15.10\\ (0.001***)\end{tabular} &
  \begin{tabular}[c]{@{}c@{}}-14.03\\ (0.001***)\end{tabular} \\
\textbf{\begin{tabular}[c]{@{}l@{}}Avg. shortest \\ path length ($L$)\end{tabular}} &
  \begin{tabular}[c]{@{}c@{}}-0.21\\ (0.314)\end{tabular} &
  \begin{tabular}[c]{@{}c@{}}-0.03\\ (0.830)\end{tabular} &
  \begin{tabular}[c]{@{}c@{}}-0.01\\ (0.967)\end{tabular} \\
\textbf{Network Size ($N$)} &
  \begin{tabular}[c]{@{}c@{}}1.76\\ (0.001***)\end{tabular} &
  \begin{tabular}[c]{@{}c@{}}1.30\\ (0.006***)\end{tabular} &
  \begin{tabular}[c]{@{}c@{}}0.98\\ (0.042*)\end{tabular} \\
\textbf{\begin{tabular}[c]{@{}l@{}}Avg. clustering \\ coefficient ($C$)\end{tabular}} &
  \begin{tabular}[c]{@{}c@{}}-0.16\\ (0.504)\end{tabular} &
  - &
  \begin{tabular}[c]{@{}c@{}}0.67\\ (0.024**)\end{tabular} \\
\textbf{Density ($d$)} &
  - &
  \begin{tabular}[c]{@{}c@{}}-0.46\\ (0.093)\end{tabular} &
  \begin{tabular}[c]{@{}c@{}}-0.95\\ (0.006***)\end{tabular} \\ \hline 
\textbf{$R^2$} &
  0.163 &
  0.183 &
  0.223 \\ \hline
\end{tabular}
\label{tab:my-table}
\end{table}

\begin{comment}
\begin{table}[h!]
\caption{The regression coefficients of each of the three linear models.}
\begin{tabular}{lccc}
\hline
\textbf{} &
  \textbf{$M_1$} &
  \textbf{$M_2$} &
  \textbf{$M_3$} \\ \hline
\textbf{Intercept} &
  \begin{tabular}[c]{@{}c@{}}-15.29\\ (0.001***)\end{tabular} &
  \begin{tabular}[c]{@{}c@{}}-15.23\\ (0.001***)\end{tabular} &
  \begin{tabular}[c]{@{}c@{}}-15.56\\ (0.001***)\end{tabular} \\
\textbf{\begin{tabular}[c]{@{}l@{}}Avg. shortest \\ path length ($L$)\end{tabular}} &
  \begin{tabular}[c]{@{}c@{}}-0.48\\ (0.004***)\end{tabular} &
  \begin{tabular}[c]{@{}c@{}}-0.37\\ (0.031*)\end{tabular} &
  \begin{tabular}[c]{@{}c@{}}-0.40\\ (0.017***)\end{tabular} \\
\textbf{Network Size ($N$)} &
  \begin{tabular}[c]{@{}c@{}}2.31\\ (0.001***)\end{tabular} &
  \begin{tabular}[c]{@{}c@{}}1.81\\ (0.001***)\end{tabular} &
  \begin{tabular}[c]{@{}c@{}}1.65\\ (0.001***)\end{tabular} \\
\textbf{\begin{tabular}[c]{@{}l@{}}Avg. clustering \\ coefficient ($C$)\end{tabular}} &
  \begin{tabular}[c]{@{}c@{}}0.21\\ (0.315)\end{tabular} &
  - &
  \begin{tabular}[c]{@{}c@{}}0.62\\ (0.025**)\end{tabular} \\
\textbf{Density ($d$)} &
  - &
  \begin{tabular}[c]{@{}c@{}}-0.27\\ (0.375)\end{tabular} &
  \begin{tabular}[c]{@{}c@{}}-0.93\\ (0.028*)\end{tabular} \\ \hline 
\textbf{$R^2$} &
  0.32 &
  0.340 &
  0.37 \\ \hline
\end{tabular}
\label{tab:my-table}
\end{table}
\end{comment}

The results provided in Table \ref{tab:my-table} show the regression coefficients of the three models. In all three of the models, sentiment diversity increases as a function of the size of a network however the average shortest path length is not a significant covariate. In $M_1$ and $M_2$, neither the average clustering coefficient nor the density adds explanatory power beyond that of just the network size. In the third model, however, we see compounding effects in which increasing network density reduces diversity while higher clustering increases it. A possible interpretation of this result is that greater interconnectedness, in general, captures the known effects of intermedia agenda-setting in which news sources tend to mimic one another and mutually reinforce each other. However, the formation of local subgroups in the media contrasts this effect and allows different groups to express different sentiment on the same issue and therefore reduce the global consensus.

\section{Conclusion}
\label{conclusion}

In this paper, we presented a method to operationalise intermedia agenda-setting by identifying networks of relationships in the media that capture how news sources influence one another to adopt particular sentiment on issues.

We used network analysis to explore the propagation of sentiment about a topic in the media. In doing so, we revealed characteristics of online news dissemination and the sentiment adoption process of the media. We found that, consistent with the expectations of intermedia agenda-setting theory, there is identifiable influence among the news media, both where sources lead/lag the herd, as well as on a peer-to-peer level. While those who lead the consensus sentiment vary from topic to topic, we found that elite news sources, such as the \textit{New York Times, Washington Post} etc., are more likely to be identified as a bellwether than those who are not elite. Moreover, we found that the larger, well established news sources also tend to have higher centrality in the networks that directly measure peer-to-peer influence. This observation contrasts the findings of other recent studies that explore the level of influence maintained by different actors in the media. A study by Vargo and Guo \cite{vargo_networks_2017} on the influence between types of media concluded that, in response to a shifting audience, the large elite media have become attentive to the agendas of smaller online news sources. Our results suggest that, while the elite media sources can be attentive to others, they do still have comparably more power. 

When compared to a null model, we identified that some topics covered by the media form network structures in which influence is highly centralised to a a small subset of news sources. However this effect was only observed on 31.5\% of the available topics, so we are unable to definitively state whether or not such effects are systematic. We did, however, find pervasive evidence of small-world characteristics between interacting news sources, and that this phenomenon is pervasive across topics. Previous work on the network structure of news dissemination has also found small-world properties when investigating the propagation of breaking news \cite{liu_breaking_2016}. While such work demonstrated that breaking news has the ability to propagate quickly, by focusing on the associated sentiment of prolonged topics, the observed small-worldness in our results indicates that changes in the framing of a story are also subject to being rapidly disseminated. Finally, by comparing descriptive networks statistics to the dynamics of how the media discusses topics, we found evidence that intermedia influence can be a positive factor in presenting diverse perspectives. In particular, we saw that while greater degrees of interconnectedness (in the form of more densely connected networks) were associated with reduced diversity in the overall sentiment that the media presents on an issue, this effect was diminished for topics in which the media forms clusters of influence.

As the first study of its kind to look at the impact that the structure of an influence network has on how perspectives are adopted in the media, there are several questions that we hope to answer in future work. Notably, we would like to include entity-specific information in our model, primarily whether the results change when explicitly controlling for the the size of a news source, whether there are partisan effects or differences between news agencies and publishers, and how the opinion dynamics vary differently within communities versus between them. We also believe that there is more to be understood of the nature of the influence through further studying the impact of the magnitudes and duration of the lead-lag relationships.

%%%%%%%%%%%%%%%%%%%%%%%%%%%%%%%%%%%%%%%%%%%%%%
%%                                          %%
%% Backmatter begins here                   %%
%%                                          %%
%%%%%%%%%%%%%%%%%%%%%%%%%%%%%%%%%%%%%%%%%%%%%%

\begin{backmatter}

%\section*{Competing interests}
%  The authors declare that they have no competing interests.

\section*{Author's contributions}
SS designed the research and conducted the experiments. GL and RES supervised the research. All authors wrote the manuscript.

\section*{Acknowledgements}
We thank Nora Ptakauskaite, Simone Righi, and David Tuckett for helpful comments on preliminary versions of this manuscript. GL acknowledges support from an EPSRC Early Career Fellowship (Grant No. EP/N006062/1).

%%  After submission of the .TEX file,                     %%
%%  you will be prompted to submit your .BBL file.         %%

% if your bibliography is in bibtex format, use those commands:
\bibliographystyle{bmc-mathphys} % Style BST file (bmc-mathphys, vancouver, spbasic).
\bibliography{bibliography}      % Bibliography file (usually '*.bib' )
% for author-year bibliography (bmc-mathphys or spbasic)
% a) write to bib file (bmc-mathphys only)
% @settings{label, options="nameyear"}

% or include bibliography directly:
% \begin{thebibliography}
% \bibitem{b1}
% \end{thebibliography}

%%%%%%%%%%%%%%%%%%%%%%%%%%%%%%%%%%%
%%                               %%
%% Figures                       %%
%%                               %%
%% NB: this is for captions and  %%
%% Titles. All graphics must be  %%
%% submitted separately and NOT  %%
%% included in the Tex document  %%
%%                               %%
%%%%%%%%%%%%%%%%%%%%%%%%%%%%%%%%%%%
\newpage
\appendix
\section*{Appendix A}
\label{appendix:A}
\begin{figure}[!htb]
  \centering
  \includegraphics[width=0.96\textwidth]{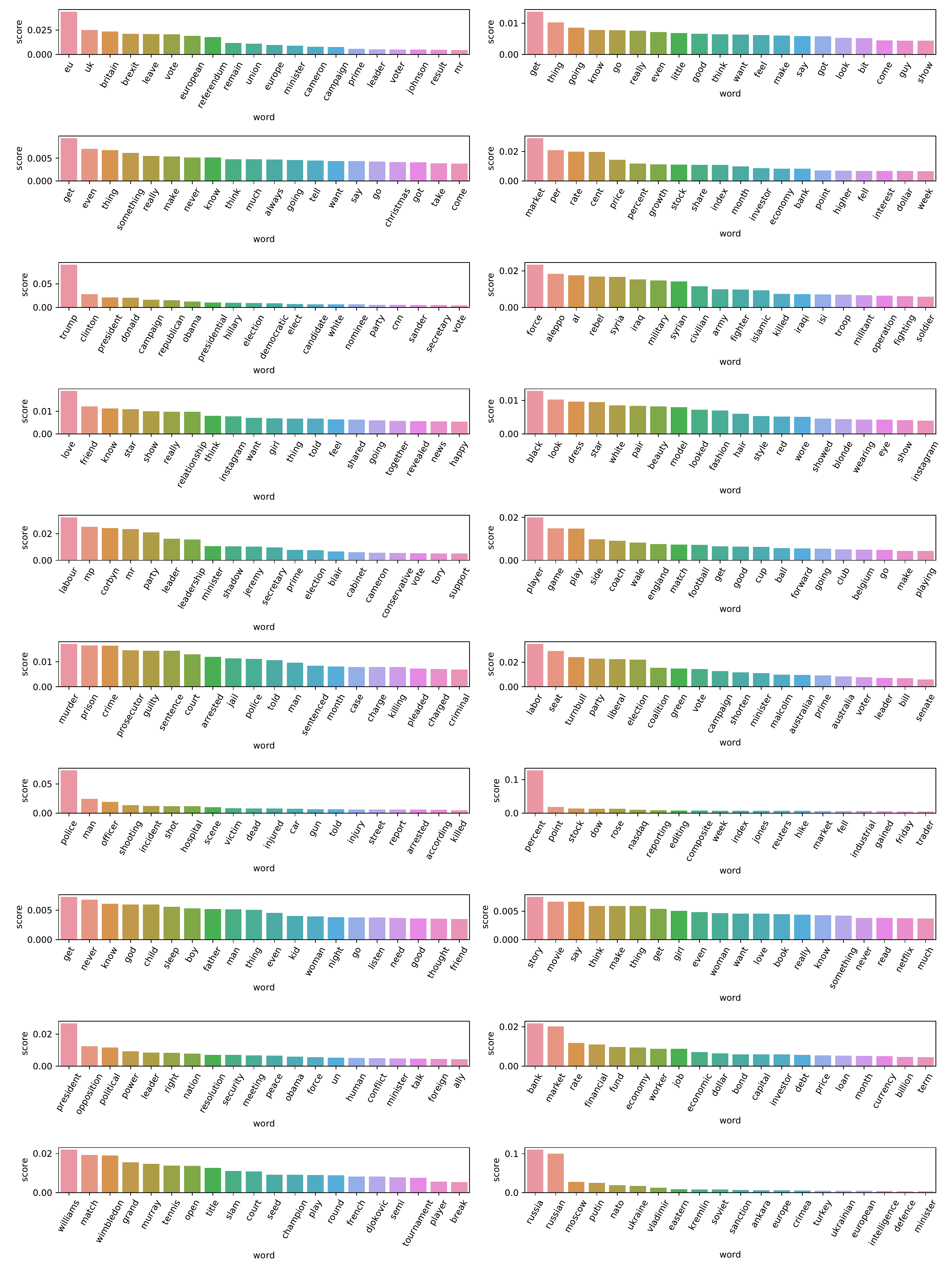}
  \caption{The inferred topics with the 20 highest coherence values}
  \label{top_topics}
\end{figure}

%%
%% Do not use \listoffigures as most will included as separate files

%%%%%%%%%%%%%%%%%%%%%%%%%%%%%%%%%%%
%%                               %%
%% Additional Files              %%
%%                               %%
%%%%%%%%%%%%%%%%%%%%%%%%%%%%%%%%%%%
\newpage

%\section*{Additional Files}
%  \subsection*{Additional file 1 --- Sample additional file title}
%    Additional file descriptions text (including details of how to
%    view the file, if it is in a non-standard format or the file extension).  This might
%    refer to a multi-page table or a figure.

%  \subsection*{Additional file 2 --- Sample additional file title}
%    Additional file descriptions text.

\end{backmatter}
\end{document}